\documentclass[11pt]{article}
\usepackage{amsmath,amsfonts}
\usepackage{tikz}
\usetikzlibrary{trees,er,snakes,shapes,mindmap}
\allowdisplaybreaks
\def \beq  {\begin{equation}}
\def \eeq  {\end{equation}}
\def \beqar {\begin{eqnarray}}
\def \eeqar {\end{eqnarray}}
\def\sqr#1#2{{\vcenter{\vbox{\hrule height.#2pt
\hbox{\vrule width.#2pt height#1pt \kern#1pt
\vrule width.#2pt}\hrule height.#2pt}}}}

\def\vf {{\varphi}}

\def\Tr {{\rm Tr}}

\def\vp {{\vec p}}

\def\vf {{\varphi}}

\def\del {\partial}

\def\a {\alpha}
\def\b {\beta}

\def\D {{\cal D}}

\def\A {{\cal A}}

\def\O {{\cal O}}

\def\half{\textstyle{1\over 2}}

\begin{document}
\fontfamily{cmr}\fontsize{11pt}{15pt}\selectfont
\def \CMP {{Commun. Math. Phys.}}
\def \PRL {{Phys. Rev. Lett.}}
\def \PL {{Phys. Lett.}}
\def \NPBProc {{Nucl. Phys. B (Proc. Suppl.)}}
\def \NP {{Nucl. Phys.}}
\def \RMP {{Rev. Mod. Phys.}}
\def \JGP {{J. Geom. Phys.}}
\def \CQG {{Class. Quant. Grav.}}
\def \MPL {{Mod. Phys. Lett.}}
\def \IJMP {{ Int. J. Mod. Phys.}}
\def \JHEP {{JHEP}}
\def \PR {{Phys. Rev.}}
\def \JMP {{J. Math. Phys.}}
\def \GRG{{Gen. Rel. Grav.}}
\begin{titlepage}
\null\vspace{-62pt} \pagestyle{empty}
\begin{center}
\rightline{CCNY-HEP-14/3}
\rightline{June 2014}
\vspace{1truein} {\Large\bfseries
Relativistic Particle and Relativistic Fluids: Magnetic 
}\\
\vspace{6pt}
\vskip .1in
{\Large \bfseries  Moment and
Spin-Orbit Interactions}\\
\vskip .1in
{\Large\bfseries ~}\\
{\large\sc Dimitra Karabali$^a$} and
 {\large\sc V.P. Nair$^b$}\\
\vskip .2in
{\itshape $^a$Department of Physics and Astronomy\\
Lehman College of the CUNY\\
Bronx, NY 10468}\\
\vskip .1in
{\itshape $^b$Physics Department\\
City College of the CUNY\\
New York, NY 10031}\\
\vskip .1in
\begin{tabular}{r l}
E-mail:&{\fontfamily{cmtt}\fontsize{11pt}{15pt}\selectfont dimitra.karabali@lehman.cuny.edu}\\
&{\fontfamily{cmtt}\fontsize{11pt}{15pt}\selectfont vpn@sci.ccny.cuny.edu}
\end{tabular}

\vspace{.8in}
\centerline{\large\bf Abstract}
\end{center}
We consider relativistic charged particle dynamics and relativistic  magnetohydrodynamics
using symplectic structures and actions given in terms of co-adjoint orbits of the Poincar\'e group.
The particle case is meant to clarify some points such as how minimal coupling
(as defined in text) leads to a gyromagnetic ratio of $2$, and to set the stage for fluid dynamics.
The general group-theoretic framework is further explained and is then used to set up Abelian magnetohydrodynamics including spin effects. An interesting new physical effect is precession of
spin density induced by gradients in pressure and energy density. The Euler equation
(the dynamics of the velocity field) is also modified by gradients of the spin density.

\end{titlepage}

\pagestyle{plain} \setcounter{page}{2}
\section{Introduction}
The study of the relativistic point-particle is a story as old as the theory of relativity itself.
This long history might suggest that there would be very little new one can say on this matter.
Nevertheless, over the years, new approaches and clarifications have been obtained
\cite{general}.
In the quantum theory, a point-particle is defined as a unitary irreducible representation
(UIR) of the Poincar\'e group. Thus, even classically, a description which highlights this connection is interesting.
The development of geometric quantization furnished the basic framework for carrying this out.
The use of symplectic forms defined on co-adjoint orbits of the group led to new Lagrangian descriptions incorporating spin and, in the case of charged particles, magnetic moments and
spin-orbit couplings \cite{SkSt}.
A Lagrangian derivation of the Bargmann-Michel-Telegdi equation for spin precession \cite{BMT} was another result of such descriptions \cite{{general}, {SkSt}, {Dixon}}.

Our return to this old problem is motivated by the potential to generalize it to
fluid mechanics. A charged fluid would obviously be described by magnetohydrodynamics.
An action-based canonical approach to magnetohydrodynamics
does exist, but we can go further and ask the question: How do we incorporate the effects
of magnetic moments and spin-orbit couplings in magnetohydrodynamics?

The description of fluids in terms of group theory has been developed over the last few years
\cite{bistro}; see also \cite{review}.
Fluids for which the constituents carry spin or internal (Abelian or nonabelian) symmetries can be described using the Lorentz or internal symmetry groups.
One particular advantage of this is the straightforward symmetry based inclusion of anomalies, which has led to formulae for the chiral magnetic and chiral vorticity effects \cite{{NRR}, {CNT}}. 
(The chiral magnetic effect, although not from a group-theory point of view,
 was first discussed in
\cite{kharzeev}.
The effects of anomalies in fluids have been analyzed in
\cite{{son+}, {landsteiner}, {Lin+}}.)
However, even though spin is naturally included in this framework
via the Lorentz group, the extension of this to the full Poincar\'e group
had some subtleties and nuances related to the fact that individual particle positions have no meaning from a fluid point of view \cite{capasso}.
We sort out these issues in this paper, as they have not been fully clarified in previous work.
Finally, in working out the fluid connection, we realized that some aspects of the role of the symplectic structure for the orbit of the Poincar\'e group for charged particles were also not entirely clear in the literature. This is another issue that is addressed in this paper.

To summarize, we will start, in section 2, by considering the symplectic structure for charged point-particles defined purely group theoretically in terms of the Poincar\'e group.
The equations of motion will be shown to lead naturally to magnetic moment and spin-orbit interactions (section 3).
At the minimal level of gauging or introducing the electromagnetic field, the Lorentz force will fix the gyromagnetic ratio to be 2, just as it is for spinning particles in single particle quantum mechanics. The Hamiltonian for proper time evolution, which is worked out
in section 4, explicitly shows this result.
This is the $(3+1)$-dimensional analog (with its own complications) of the similar situation in
$(2+1)$ dimensions \cite{{anyon-us}, {cortes}, {horvathy}}. 
Nonminimal coupling can be introduced to account for the anomalous magnetic moment, just as in $(2+1)$ dimensions \cite{horvathy}. (We should caution that the word ``minimal" is used
with somewhat different meanings here, in \cite{SkSt} and in \cite{horvathy}. Also, one might entertain variants of
the Lorentz force. But the following conditional statement is valid:
The Lorentz force 
 in terms of the coordinate in the symplectic form, the latter being given purely by the group structure, leads to $g = 2$.) In the $3+1$ case, an arbitrary magnetic moment has been included in the proposed Lagrangians in \cite{SkSt}; however, the special role of $g = 2$ is not manifest, or, at least,
 not highlighted
 in this approach.
 We also show how a modification of the symplectic structure can accommodate $g \neq 2$.
 The proper time Hamiltonian should be zero as a constraint, for example, on states upon quantization. This is the single particle wave equation.
Writing this in terms of mutually commuting coordinates, which we do in section 5, again shows explicitly the magnetic
moment and spin-orbit interactions. This is also very much a replay of the similar situation in
$2+1$ dimensions.
 
 Starting from the symplectic structure for the point-particle and following Lagrange's method of
 obtaining fluid dynamics from particle dynamics, we obtain the required fluid action in section 6.
 The variation of the particle coordinates, now viewed as a field, can still lead to the equations of motion,
 but incorporating nonzero pressure is awkward in this language.
 This is again related to the lack of a suitable
 fluid interpretation of particle positions alluded to before.
 For this reason, we switch to the Clebsch parametrization at this stage.
 The end result is an action for a relativistic fluid with spin, magnetic moment, spin-orbit interactions, etc.
 In section 7, we examine standard Abelian magnetohydrodynamics in some detail, but now including
 spin.
The Euler equation for the charged fluid now shows additional force terms involving the gradients of the spin density. This is not surprising by itself since it is related to the magnetic moment interaction.
What is more interesting is the precession equation for the spin density. This has, in addition to the usual
term proportional to the external field, terms involving gradients of the pressure and energy density.

A concluding short discussion summarizes the new results in this paper.

Before concluding this section, we also mention that
there has been some recent work on the
 use of the symplectic form derived from the Poincare group to discuss
 Dirac and Weyl particles \cite{Stone}.
 While our focus has been on fluids, and hence not directly along these lines,
 clearly there is some resonance with
 our work.
 
 \section{The symplectic form}
 
 We begin by recalling the essence of the co-adjoint orbit method.
 If $g$ denotes a general element of a Lie group $G$, in a particular representation,
 the action is
 \beq
 S = i  \sum_\alpha w_\alpha \, \int d\tau ~ \Tr ( h_\alpha \, g^{-1} {\dot g} ),
 \hskip .3in {\dot g} = {d g \over d \tau}
 \label{1}
 \eeq
 where $h_\alpha$ give a basis of the diagonal generators of the Lie algebra (the Cartan subalgebra)
 and $w_\alpha$ are a set of numbers. We are envisaging a matrix representation, say, the fundamental representation, with $\Tr (h_\alpha \, h_\beta ) = \delta_{\alpha\beta}$.
  The basic theorem is that the quantization of this action leads to a Hilbert space
 which carries a unitary irreducible representation of $G$, this UIR being specified by the highest weight $(w_1, w_2, \cdots, w_r)$, $r$ being the rank of the group, which is also the range of summation for
 $\alpha$. The canonical one-form associated to (\ref{1}) is evidently
 \beq
 \A = i \sum_\alpha w_\alpha \, \Tr ( h_\alpha \, g^{-1} { d g} )
 \label{2}
 \eeq
 Under transformations $g \rightarrow g \, \exp (-i h _\alpha \vf_\alpha )$, we find
 $\A \rightarrow \A + d f $, $f = \sum w_\alpha \vf_\alpha$.
 Thus the symplectic two-form $\Omega = d \A$ is defined on $G/T$, $T$ being the maximal torus.
 Further, the transformation $\A \rightarrow \A + df$ shows that in the quantum theory, where wave functions transform as $e^{iS}$, there will be restrictions or quantization conditions on $w_\alpha$,
 these being the appropriate conditions for $(w_1, w_2, \cdots, w_r)$ to qualify as the highest weight
 of a UIR.
 
In extending this directly to the Poincar\'e group, first of all, there is a small technical difficulty
due to the lack of a matrix representation with a well-defined trace. One can use infinite dimensional representations, but the notion of the trace has to be carefully defined.  This can be done with a suitable regularization or a cut-off on integrals, but a  simpler solution is to use a finite-dimensional
 representation of the de Sitter group, obtaining the Poincar\'e group as a contraction. Thus we will use 
 \beqar
 P_\mu &= &\gamma_\mu / r_0 \nonumber \\
 J_{\mu\nu} &=&  \gamma_{\mu\nu}= (i/4) [ \gamma_\mu , \gamma_\nu ]
 \label{2prime}
 \eeqar
 as the representation of the de Sitter algebra, $\gamma_\mu$ being the standard $4\times 4$ Dirac matrices.
 The parameter $r_0$ which is the radius of curvature of the de Sitter
 space, can be taken to be very large to recover the Poincar\'e limit. A general group
 element is of the form
 \beq
 g = \exp ( i P_\mu \, x^\mu ) \, \Lambda
 \label{3}
 \eeq
where $\Lambda$ is a Lorentz transformation (generated by $J_{\mu\nu}$). The group has rank equal to $2$ and, as the diagonal generators, we will choose $P_0$ and $J_{12}$. The corresponding weights will be mass and spin. Our philosophy here will not be to write down actions {\it a priori}, that will come later, but to start with $\Omega$ and work out a Hamiltonian for $\tau$-evolution.
The trajectories should be invariant under reparametrizations of $\tau$, thus the Hamiltonian for $\tau$-evolution is the generator of a gauge symmetry.
We must thus set $H \approx 0$ on states in the quantum theory. This will become the wave equation for single particle states.
This was the approach followed in \cite{anyon-us} to obtain wave equations for anyons.
In this approach, there is no fixed mass, that arises from the constraint
$H \approx 0$. Rather, we should use 
$\sqrt{p^2}$ in place of the mass.
Finally, $\Lambda$ can be parametrized in terms of the boost operator connecting the rest frame to an arbitrary frame of momentum $p_\mu$. It is thus given by
\beqar
\Lambda &=& B (p)\, R \nonumber\\
B(p)&=& {1\over \sqrt{2 m (p_0 +m)}} \left[
\begin{matrix} p_0 +m & {\vec\sigma}\cdot\vp \\
{\vec\sigma}\cdot\vp & p_0 +m \\
\end{matrix}
\right]
\label{4}
\eeqar
where $m$ is a shorthand notation for $\sqrt{p^2}$. $R$ is a pure spatial rotation matrix
generated by $J_{12}, \, J_{23}, \, J_{31}$.
The canonical one-form in our case is thus given by
\beq
\A = i r_0^2 \sqrt{p^2} \, \Tr \left( {\gamma_0 \over r_0} g^{-1} \, dg \right)
+ i ~s~ \Tr ( J_{12} g^{-1} \, dg )
\label{5}
\eeq
where $s$ denotes the spin of the particle.
Using $B \gamma_0 B^{-1} = \gamma^\alpha \, p_\alpha/ \sqrt{p^2}$ and taking
$r_0 \rightarrow \infty$, we find, for the Poincar\'e group,
\beq
\A = - p_\mu \, dx^\mu +i {s \over 2} \Tr (\Sigma_3 \Lambda^{-1} \, d\Lambda ),
\hskip .3in \Sigma_a = \left[ \begin{matrix} \sigma_a &0 \\  0&\sigma_a\\ \end{matrix} \right]
\label{6}
\eeq
Upon using (\ref{4}), this simplifies as
\beqar
\A &=& - p_\mu \, dx^\mu + i {s \over 2} \Tr (\Sigma_3 \, R^{-1} \, D R )\nonumber\\
DR &=& dR ~+~ C \, R\nonumber\\
C&=& - i \left( {\Sigma_a \over 2}\right) {\epsilon_{abc} p_b \, dp_c \over m (p_0 +m)},
\hskip .2in R = \exp \left[ i \left( {\Sigma_a \over 2}\right) \theta^a\right]
\label{7}
\eeqar
We will see that $R$ describes the spin degrees of freedom, with $2s$ quantized
as an integer, so that the wave functions are appropriately single-valued or double-valued for fermions.

The expression for $\A$ does not seem to have manifest Lorentz invariance
(with $p_\mu$ taken as a $4$-vector), but this is just as it should be. Under a Lorentz transformation
$p_\mu \rightarrow S_\mu^{~\nu} \, p_\nu$, we have $B (Sp) = {\hat S} \, B(p) \, R_W$
where $\hat S$ is the Dirac spinor version of $S$ and $R_W$ is the Wigner rotation
corresponding to the parameters in $S$.
With $R \rightarrow R^{-1}_W\, R$ accompanying $S$, we find
\beq
B(p) \, R \rightarrow B (S p) \, R^{-1}_W R = {\hat S} \, (B R)
\label{8}
\eeq
giving the required covariance property. Defining the unit vector $N_a$ by
\beq
N_a \, \Sigma_a = R \, \Sigma_3 \, R^{-1}
\label{9}
\eeq
the symplectic potential can be simplified as
\beq
\A = - p_\mu \, dx^\mu  
+ s\, { \epsilon_{abc} N_a\, p_b\, dp_c \over m\, (p_0 +m)}
+ i {s \over 2}\,  \Tr (\Sigma_3 R^{-1}\, dR )
\label{10}
\eeq
It is easy enough to see that the transformations $R \rightarrow \left[ 1- i {\vec \Sigma}\cdot {\vec \epsilon}/2\right] \, R$ are generated by $s \, N_a$ in the rest frame identifying
$S_a = s\, N_a$ as the spin vector in the rest frame. We can also work out $\Omega^{(0)}
= d \A$ as
\beqar
\Omega^{(0)} &=& dx^\mu \, dp_\mu  + {S^{\mu\nu} dp_\mu \, dp_\nu \over 2\, m^2}+ s { \epsilon_{abc} dN_a\, p_b\, dp_c \over m\, (p_0 +m)}
\nonumber\\
&& \hskip .2in
- {s \over 2} (\vec {N} \cdot\vp)  {\epsilon_{abc} p_a \, dp_b \, dp_c \over m^2 \, (p_0 +m)^2}- {s \over 2} \epsilon_{abc} \, N_a\, dN_b \, dN_c
\label{11}
\eeqar
where $S^{\mu\nu}$ is the canonical generator of the Lorentz transformation $\Lambda \rightarrow
( 1- i \omega^{\mu\nu} \, J_{\mu\nu} )\, \Lambda$ given by
\beq
S_{\mu\nu} = {s \over 2} \Tr (\Sigma_3 \Lambda^{-1} J_{\mu\nu} \Lambda)
\label{11prime}
\eeq
This can be explicitly written in terms of the spin vectors as
\beqar
S_{0i} &=& - \epsilon_{ijk} {p_j S_k \over m}\nonumber\\
S_{ij}&=& \epsilon_{ijk} \left[ {p_0 S_k \over m} - {{\vec S} \cdot \vp \,\, p_k \over m (p_0 +m)}
\right]
\label{12}
\eeqar
Evidently
\beq
S^{\mu\nu}\, p_\nu = 0
\label{13}
\eeq
Since $N^2 = 1$, $\epsilon_{abc} N_a \, dN_b \, dN_c$ is proportional to
the volume (area) of the two-sphere defined by $N_a$. The requirement
that the integral of $\Omega$ on any closed two-surface should be $2 \pi$ times an integer
shows that $2s$ must be an integer upon quantization.

We now turn to the introduction of the electromagnetic field. The usual minimal prescription amounts to adding $\int d\tau~ e A_\mu {\dot x}^\mu $ to the action. This is equivalent to adding 
$e A_\mu dx^\mu$ to $\A$ so that
\beq
\Omega = \Omega^{(0)} ~+ ~ {e \over 2} \, F_{\mu\nu} \, dx^\mu \, dx^\nu
\label{14}
\eeq
We will refer to this as the ``minimal prescription". (As mentioned in the introduction, there are some
variations in the meaning attributed to the word ``minimal" in the literature.)

\section{The equations of motion}

Our next step is to consider the equations of motion, which will determine the 
Hamiltonian for $\tau$-evolution.
We consider uniform fields $F_{\mu\nu}$ viewed as a good approximation to slowly varying fields.
The `minimal' equations of motion will be taken to be
 the usual Lorentz force equations
 \beqar
 {d x^\mu \over d \tau}&=& {p^\mu \over M} \nonumber\\
 {d p^\mu \over d \tau} &=& - e\, F^{\mu\nu} {d x_\nu \over d \tau} ~+~ {\cal O} (\del F )
 \label{18}
 \eeqar
Thus, our definition of minimal coupling and Lorentz force amounts to the $\Omega$ given in
 (\ref{11}), (\ref{14}) and the equations of motion given in (\ref{18}).

It is useful to consider some arguments motivating our labeling of the trio
 (\ref{11}), (\ref{14}), (\ref{18}) as minimal coupling, even though it is not strictly needed
 in the flow of logic taking us to the Hamiltonian.
The general form of the equations of motion consistent with the Lorentz force is given by
\beqar
{d x^\mu \over d \tau} &=& c \, \left( {p^\mu \over M} + L^\mu \right)
\nonumber\\
{d \over d\tau} \left[ c \, (p^\mu + M \, L^\mu ) \right]
&=& - e F^{\mu \nu} ~c\, \left( {p_\nu \over M} + L_\nu \right) ~+~
{\cal O} (\del F )
\label{15}
\eeqar
where $c$ is a constant of motion, $M$ is a constant mass and $L^\mu$ is an arbitrary function of $S_a$, $p_\mu$ and $F_{\mu\nu}$. We can see that by eliminating $p^\mu$ in favor of the velocity
we get the usual Lorentz force for the $\tau$-evolution of $x^\mu$.
Now, since $c$ is a constant of motion, we can write $\tau = \lambda /c$, so that equations
of motion may be reduced to the form
\beqar
{d x^\mu \over d \lambda} &=& \left( {p^\mu \over M} + L^\mu \right)
\nonumber\\
M {d^2 x^\mu \over d\lambda^2} &=& {d \over d\lambda} ( p^\mu + M \, L^\mu)
= -e F^{\mu\nu} \, {dx_\nu \over d\lambda}  ~+~ {\cal O} (\del F)
\label{16}
\eeqar
This means that, since we have reparametrization invariance for $\tau$ (once we set its evolution operator to zero), we may, 
without loss of generality, set $c=1$. (The freedom of such a function $c$ was noted in 
\cite{horvathy}, where the authors also noted that it amounted to just reparametrization of $\tau$.) The story with $L^\mu$ in (\ref{16}) is more involved.
 The $4$-velocity in terms of $x^\mu$ is
 \beq
 u^\mu = {d x^\mu \over d\lambda} {1\over \sqrt{ {dx^\alpha \over d\lambda}{dx_\alpha \over d\lambda}}}
 \label{17}
 \eeq
(We write in this way to make it independent of the parameter $\lambda$.) 
However, the Lorentz boost transformation $B(p)$
takes us from the rest frame to one which is moving
 with $4$-velocity $p^\mu /\sqrt{p^2}$, as is clear from the explicit formula for $B(p)$.
Thus, {\it a priori}, there are two velocities which enter the description.
Since a moving particle can be brought to rest by a suitable boost defined by
$B(p)$, we should expect these two velocities to be the same, namely, that
$u^\mu$ (as defined in (\ref{17})) to be
$p^\mu /\sqrt{p^2}$.
This would be possible for all values of
 $F_{\mu\nu}$ and $p_\mu$ only if $L^\mu = 0$. ($L^\mu \sim p^\mu$ is possible
 but that is equivalent to having $c$, which we have already discussed.)
 In short, if
the symplectic structure is entirely defined by the Poincar\'e group,
boost transformations (with the velocity parameter occurring in $B(p)$) should
implement the transformation from the rest frame to the 
comoving frame of the particle.
Incorporating this feature in 
 (\ref{16}), we get the equations of motion (\ref{18}).
 This motivates our qualification of (\ref{18}) as minimal.
 
So what is nonminimal? Our expression for $\Omega^{(0)}$ is given by Poincar\'e symmetry.
The gauging is done by adding 
the term $e A_\mu dx^\mu$ in $\A$, to the canonical one-form;
this
is basically singled out as the leading coupling for slowly varying fields by gauge invariance.
Going beyond what we termed minimal would include
additional terms in $\Omega^{(0)}$, possible changes to the equations of motion themselves, etc.
For example, one could envision corrections to the equations of motion involving powers of $F$, even when we ignore gradients of the fields.
(We may, however, note that corrections to the equations of motion with higher powers of
the field strength 
must involve powers of $F_{\mu\nu}/p^2$ or $F_{\mu\nu}/M^2$ for dimensional reasons.
Thus they are significant only at high field strengths, of magnitudes needed for pair production. Single particle dynamics may not be adequate for such field strengths anyway.)

 Our analysis is for the minimal case, except for the anomalous magnetic moment discussed briefly later.

\section{The Hamiltonian}

We are now ready to obtain the Hamiltonian for $\tau$-evolution. This can be done in terms of the $\tau$-evolution vector field given by (\ref{18}) or by computing the Poisson brackets. Either way, we need Hamiltonian vector fields $V_f$, corresponding to a function $f$, defined by
\beq
V_f \rfloor \, \Omega = - d \, f
\label{19}
\eeq
 where $V \rfloor \Omega$ denotes the interior contraction of $V$ with
 $\Omega$ given by
 \beqar
 V \rfloor \Omega &=&  V^\mu \, \Omega_{\mu\nu} \, d \xi^\nu\nonumber\\
 \Omega &=& {1\over 2} \, \Omega_{\mu\nu} \, d \xi^\mu \, d\xi^\nu , \hskip .3in 
 V= V^\mu {\del \over \del \xi^\mu}
 \label{20}
 \eeqar
$\xi^\mu$ are phase space coordinates given by $x^\mu$, $p_\mu$ and the two coordinates on the two-sphere defined by $N_a$. First define a vector ${\upsilon}^\mu$, ${\upsilon}^0 =0$
 and ${\upsilon}^i$ by its action on $N_a$ as
 \beq
 {\upsilon}^i \rfloor d N_a = {N_i p_a - \delta_{ia} N \cdot \vp \over m (p_0 +m)},
 \hskip .3in m = \sqrt{p^2}
 \label{21}
 \eeq
 Although tedious, it is straightforward to verify that the vector fields corresponding to
 $p_\alpha$ and $x^\mu$ are
 \beqar
 (V_p)_\alpha &=& - (M^{-1})_\alpha^{~\mu} \left[ {\del \over \del x^\mu}
 + e F_{\mu\nu} \, Q^\nu \right]
 \nonumber\\
 (V_x)^\mu &=& \left[ \delta^\mu_{~\nu} - {S^{\mu \alpha} (M^{-1})_\alpha^{~\lambda} (e F_{\lambda \nu})  \over m^2} \right] \, Q^\nu - {S^{\mu \alpha} (M^{-1})_\alpha^{~\lambda} \over m^2}
 {\del \over \del x^\lambda}
 \label{22}
 \eeqar
 where
 \beq
 M_\mu^{~\alpha} = \delta _{\mu}^{\alpha} + {{e F_{\mu\nu}S^{\nu\alpha}} \over {m^2}}, \hskip .5in
 Q^\mu = {\del \over \del p_\mu} + {\upsilon}^\mu 
 \label{23}
 \eeq
 One can also verify that
 \beq
 Q^\mu \, S^{\alpha \beta} = {S^{\mu \alpha} p^\beta - S^{\mu \beta} p^\alpha \over m^2}
 \label{24}
 \eeq
 The Poisson bracket of two functions $f$ and $g$ is given by
 \beq
 \{f,~g \} = -(V_f \rfloor V_g \rfloor \Omega) = ( V_f \rfloor dg ) 
 \label{25}
 \eeq
 The basic bracket relations we need are thus given by
 \beqar
 \{x^{\mu},~ x^{\nu} \}& =& - {{K^{\mu \nu} } \over {m^2}}  
  \label{26}\\
  \{x^{\mu},~ p_{\nu} \} &=& \delta^{\mu}_{~\nu} - {{K^{\mu \a}(e F_{\a\nu})} \over {m^2}}  
  \label{27}\\
  \{p_{\mu},~ p_{\nu} \} &=& -\Bigl[ e F_{\mu\nu}  - {(e F_{\mu\a}){K^{ \a\b}(e F_{\b\nu})} \over {m^2}}\Bigr] 
  \label{28}\\
 \{x^{\mu},~ S^{\a\b} \} &=& {{(p^{\a}K^{\b\mu}-p^{\b}K^{\a\mu})} \over {m^2}} \\ 
 \{ p_{\mu},~ S^{\a\b} \} &=& {{p^{\a}(eF_{\mu\nu})K^{\nu\b}-p^{\b}(e F_{\mu\nu})K^{\nu\a}} \over {m^2}} 
 \label{29}\\
 K^{\mu\nu} &= &S^{\mu\alpha} (M^{-1})_{\alpha}^{\nu} \nonumber \\
 \label{30}
 &=& S^{\mu\nu} - S^{\mu\a}{{eF_{\a\b}} \over m^2} S^{\b\nu} + \cdots 
 \label{31}
 \eeqar
Since $S^{\a\b}p_{\b}=0$ and $H$ is to be made out of invariants, we can take it to be of the form
\beq
H= {{p^2} \over {2M}} + { 1 \over {2M}} ({g \over 2}) e F_{\a\b} S^{\a\b} + {\rm constant} + \O (F^2, \del F)
\label{32}
\eeq
We will see in a moment that it is adequate to take $g$ to be a constant. The PB's show that 
\beqar
{{dx^{\mu}} \over {d\tau}} &= &{p^{\mu} \over M} + ( {g \over 2} -1) {{K^{\mu\a} eF_{\a\b} p^{\b}} \over {M m^2}} \\
{{dp_{\mu}} \over {d\tau}} &=& -{{eF^{\mu\nu} p^{\nu}} \over M} - ( {g \over 2} -1) {e F_{\mu\nu} {K^{\nu\a} eF_{\a\b} p^{\b}} \over {M m^2}} 
\label{34}
\eeqar
In addition to the desirable Lorentz force terms in (\ref{18}), these equations have extra terms depending on $KF$. There are two sources for these. The term proportional to $g/2$ arises from the corresponding term in $H$ in (\ref{32}). The term with the relative coefficient $-1$ arises from the Poisson brackets (\ref{26}, \ref{29}). Because of these contributions, if we want the Lorentz force equations (\ref{18}), $g=0$ is not an option. In fact, we must have $g=2$. This is the essence of our statement in the introduction that, with the minimality conditions we have stated, the Lorentz force equations imply $g=2$. There is no surprise here; after all, it is well known that the minimal gauging of single-particle wave equations for spinning particles does lead to $g=2$. 

We now turn to the possibility of anomalous magnetic moment. For obtaining the equations of motion (\ref{18}) with $g \ne 2$, KF-dependent terms from the PB's must have a different coefficient. We must change the PB's. The minimal choice of $\Omega$ will not suffice. A bit of trial and error shows that the modification 
\beq
\Omega \rightarrow \Omega + dx^{\mu} dB_{\mu}
\label{35}
\eeq
where $B_{\mu}$ is to be determined as a function of $F_{\mu\nu}$, $S^{\mu\nu}$ and $p_{\a}$ will suffice. We will determine $B_{\mu}$ as a series in powers of $F$. If we use $\Omega + \delta \Omega$ as the symplectic form, the modified PB's are given by
\beq
\{ f,~g \} = \{ f,~g\}_{(0)} + \Bigl( V^{(0)}_f \rfloor V^{(0)}_g \rfloor \delta\Omega \Bigr) + \cdots
\label{36}
\eeq
where $V^{(0)}$ are the Hamiltonian vector fields given by $\Omega$. We find
\beqar
 \{x^{\mu},~ x^{\nu} \}& =& - {{K^{\mu \nu} } \over {m^2}}+ { 1 \over {m^2}} \Bigl[  S^{\mu\lambda}(Q^{\nu} B_{\lambda}) - S^{\nu\lambda}(Q^{\mu} B_{\lambda}) \Bigr] + \cdots
 \label{37}\\
  \{x^{\mu},~ p_{\nu} \} &=& \delta^{\mu}_{\nu} - {{K^{\mu \a}(e F_{\a\nu})} \over {m^2}} - (Q^{\mu} B_{\nu}) + \cdots
  \label{38}
  \eeqar
  The corresponding equations of motion become
  \beqar
{{dx^{\mu}} \over {d\tau}} &=& {{p^{\mu} + B^{\mu}} \over M} + ( {g \over 2} -1) {{K^{\mu\a} eF_{\a\b} p^{\b}} \over {M m^2}} - (Q^{\mu} B_{\nu}){p^{\nu} \over M} - {B^{\mu} \over M} + \cdots
\label{39}\\
{{d(p_{\mu}+B_{\mu})} \over {d\tau}} &=& -{{eF^{\mu\nu} \dot{x}^{\nu}} \over M} \label{40}
\eeqar
The choice
\beq
B_{\mu} = {{(g-2)} \over 2} {{(S^{\a\b} e F_{\a\b})} \over {2 m^2}} p^{\mu} + \O (F^2)
\label{41}
\eeq
eliminates all the unwanted terms on the right hand side of (\ref{39}), giving us the Lorentz force equations, expressed as
\beq
M \ddot{x}^{\mu} = -e \, F^{\mu\nu} \dot{x}_{\nu}
\label{42}
\eeq
The modification of the symplectic form as in (\ref{35}) corresponds to the action
\beq
S= \int {{dx^{\mu}} \over {d\tau}} (p_{\mu} + B_{\mu}) + e A_{\mu} {{dx^{\mu}} \over {d\tau}} + \cdots
\label{42prime}
\eeq 
Since $x_{\mu}$ only appears in the terms explicitly shown here, we can obtain ({\ref{40}) as an exact equation of motion. We then choose $B_{\mu}$ such that ${{dx^{\mu}} \over {d\tau}} = {{p^{\mu}+B^{\mu}} \over M}$ is also obtained as an exact equation, even though $B^{\mu}$ has to be determined in a series expansion. Thus the Lorentz force is still an exact result, up to terms involving gradients of the fields.

Even though the form of the Hamiltonian (\ref{32}) has not changed, apart from the fact that $g$ does not have to be $2$ anymore, the modification of the PB's is crucial in obtaining the correct spin-orbit interaction as will be shown in the next section.

\section{ The wave equation}

Since $\tau$ is a gauge parameter, the generator of $\tau$-evolution, namely $H$, must be set to zero. Upon quantization, the wave equation is thus given by $H \Psi =0$. Taking the constant in (\ref{32}) to be $- \mu^2 / 2M$, this becomes
\beq
\Bigl[ p^2 + {{eg} \over 2} F_{\a\b} S^{\a\b} - \mu^2 \Bigr] \Psi = 0
\label{43}
\eeq
This shows that the mass appearing in solutions of the wave equation
is $\mu$. This would also be the mass which appears in the classical equations of motion obtained starting from the wave equation and taking a classical limit. However, the
mass parameter we used for the classical 
equations of motion was $M$. There is no real contradiction here by virtue of reparametrization invariance.
Going back to (\ref{18}) or (\ref{42}), we see that $M$ can be replaced by $\mu$ by redefining $\tau \rightarrow \tau M / \mu$. Therefore, without loss of generality we can take $\mu = M$, and the wave equation is 
\beq
\Bigl[ p^2 + {{eg} \over 2} F_{\a\b} S^{\a\b} - M^2 \Bigr] \Psi = 0
\label{44}
\eeq
The orbit of the Poincar\'e group, and the corresponding $\Omega$, should be specified by the values for the set of mutually commuting observables, which are the mass and the spin. But we relaxed the condition to retain unconstrained $p_{\mu}$ in $\Omega$ by using $\sqrt{p^2}$ in place of the mass, as in (\ref{5}). The requirement (\ref{44}) is thus the reinstatement of the definition of mass. 

As is evident from (\ref{37}), in the quantum theory, $x^{\mu}$ does not commute with $x^{\nu}$. Thus, to write (\ref{44}) as a differential equation, we must first transform to a mutually commuting set of coordinates. This is equivalent to the choice of Darboux coordinates in the classical theory. 
We will show how this can be done for the simpler case of $g = 2$, where $B_\mu = 0$.
In this case, it is easily verified that the required transformation is
\beqar
x^\mu &=& q^\mu - C^\mu(k) - {e\over 2} \, F_{\lambda \alpha} \left(
{\del C^\mu \over \del k_\alpha} q^\lambda + {\del C^\alpha \over \del k_\mu} \,C^\lambda
\right) + \cdots\nonumber\\
p_\mu &=& k_\mu - {e\over 2} F_{\mu \alpha} \, q^\alpha + e F_{\mu \alpha} \, C^\alpha(k)
- {1\over 4} q^\lambda e F_{\lambda \alpha} \left( {\del C^\beta \over \del k_\alpha}
- {\del C^\alpha \over \del k_\beta}\right) \, e F_{\beta \mu} + \cdots
\label{44a}
\eeqar
where $q^\mu$, $k_\mu$ are standard canonical coordinates,
\beq
[ q^\mu , q^\nu ] = 0, \hskip .2in [q^\mu , k_\nu ] = \delta^\mu_\nu , \hskip .2in
[k_\mu , k_\nu ] = 0
\label{44b}
\eeq
Further, $C^\alpha$ in (\ref{44a}) is given by
\beq
C^\alpha = S_a \, {\epsilon_{abc}\, k_b \over \sqrt{k^2} (k_0 +\sqrt{k^2} ) } \, \delta^\alpha_c 
\label{44c}
\eeq
This corresponds to the one-form $C$ in (\ref{7}) written in terms of the commuting momenta $k$. One can verify that these definitions (\ref{44a}) reproduce
the Poisson brackets
(\ref{26}) to (\ref{29}) to the lowest nontrivial order.
In evaluating Poisson brackets with (\ref{44a}), it should be kept in mind that
\beq
[ S_a, S_b ] = - \epsilon_{abc} \, S_c
\label{44d}
\eeq
The symplectic structure (\ref{11}), specifically the term
$-(s/2) \epsilon_{abc} N_a \, dN_b \, dN_c$, gives these PB relations for $S_a$.

The Darboux coordinates (\ref{44a}) show that, in going to the quantum theory, we can represent
$p_\mu$ in terms of derivatives with respect to $q^\mu$ as
\beq
p_\mu = -i \, {\del \over \del q^\mu } + e A_\mu + e F_{\mu\alpha} \, C^\alpha + \cdots
\label{44e}
\eeq
where $A_\mu = - {\half} F_{\mu \alpha} q^\alpha$ is the vector potential
for us, since we are ignoring derivatives of $F_{\mu\alpha}$.
The wave equation (\ref{44}) thus takes the form
\beq
\left[ - ( \del_\mu + i e A_\mu +i e F_{\mu \alpha } C^\alpha + \cdots  )^2
+ {e\over 2} g\, S^{\alpha\beta}\, F_{\alpha\beta} ~-~ M^2 \right] \, \Psi = 0
\label{44f}
\eeq
One point worth emphasizing here is that while the term
$(eg /2) S^{\alpha\beta} F_{\alpha \beta}$ gives the correct magnetic moment
interaction, the spin-orbit interaction part of this term is twice what is needed.
The extra term
$ -  i e (\del_\mu F_{\mu\alpha} C^\alpha + F_{\mu\alpha} C^\alpha \del_\mu ) \Psi$
compensates for this and leads to the correct spin-orbit interaction in the wave equation. 
More specifically, if we introduce the
nonrelativistic wave function $\Psi_{NR}$ by writing
$\Psi= e^{i M q^0} \Psi_{NR}$,(\ref{44f}) can be simplified in the non relativistic limit as
\beqar
H \Psi_{NR} & \equiv & -i{\del \over {\del q^0}} \Psi_{NR} \nonumber \\
& \approx &  \Bigl[ -{{(\nabla + ie \vec{A})^2 } \over {2M}} -e A_0 - e F_{0i}C^i -{{eg} \over {4M}} (2 S^{0i}F_{0i} + S^{ij} F_{ij}) \Bigr] \Psi_{NR} \nonumber \\
& \approx &  \Bigl[ -{{(\nabla + ie \vec{A})^2 } \over {2M}} -e A_0 + {e \over {2 M^2}}  \vec{S} \cdot (\vec{k} \times \vec{E}) - {e \over M} \vec{S} \cdot \vec{B}  \Bigr] \Psi_{NR}
\label {44g}
\eeqar
where we have used the nonrelativistic limit of $S^{\mu\nu}$ in (\ref{12}) and $C^i$ in (\ref{44c}). Equation (\ref{44g}) shows the correct magnetic moment and the correct
spin-orbit interaction for $g=2$.

\section{Fluids}

As the first step in generalizing these considerations to fluids, we consider the action for a number of
particles, each described by $\A$ in ({\ref{10}). 
For the point we want to make, it suffices to
consider spinless particles. 
The action is thus given by
\beq
S= \sum_\alpha^N \int \Bigl[ p_\mu^{(\alpha )} \, {\dot x }^{\mu (\alpha)}
- f( n^{\alpha}) \Bigr]
\label{45}
\eeq
where $n^2 = p^2$.
The particles are labeled by the index $\alpha$. We are interested
in a continuum approximation where $N$ is very large
and the index $\alpha$ becomes almost a continuous variable.
Lagrange's key observation was that the initial positions of particles may be used to
label them, so that $\sum_\alpha^N \rightarrow \int d^3\alpha$. Further, the transformation
between the present positions $x_i (t)$ and the initial positions
is a diffeomorphism, so we can replace $d^3 \alpha$ by $d^3x\, J$,
$J$ being the Jacobian $\vert \del \alpha/ \del x\vert$. The action can then be
expressed as a spacetime integral.
The Jacobian $J$ can be absorbed into the definition of 
$p_\mu$. We then find
\beq
S = \int d^4x\, \left[ p_\mu (x) \, \left( {\del x^\mu \over \del \tau}\right)  - f(n )
\right]
\label{46}
\eeq
The $4$-velocity $u^\mu (x) = {\dot x}^\mu$  is the flow velocity of a stream of particles.
The Hamiltonian will depend on $p_\mu$. 

We know that such a simple generalization {\it at the level of the action}
will not suffice. Lagrange's derivation of hydrodynamics was done at the level of the equations
of motion and a derivation based on the action came much later, with the replacement of
the coarse-grained particle velocity in terms of the Clebsch variables.
Nevertheless, let us go a little further with (\ref{46}) and work out the equations of
motion by variation with respect to $x^\mu$. For this,
we write (\ref{46}) in terms of $d^3 \alpha$ again and find
\beqar
\delta S &=& \int d\tau d^3 \alpha~ p_\mu^{(\alpha )}\,
{\del \over \del \tau} (\delta x^\mu ) \nonumber\\
&=&  \int d\tau d^3 \alpha~ p_\mu^{(\alpha )}\,
u^\lambda \,{\del \over \del x^\lambda} (\delta x^\mu ) \nonumber\\
&=&  \int d^4x~ J\, p_\mu^{(\alpha )}\,
u^\lambda \,{\del \over \del x^\lambda} (\delta x^\mu ) \nonumber\\
&=& - \int d^4x~ \left[ {\del \over \del x^\lambda} \, ( u^\lambda \, p_\mu ) 
\right] \, \delta x ^\mu
\label{47}
\eeqar
where we used the fact that $\del / \del \tau$ acting on a function $g$ of
$x^\mu$ is $u^\lambda ({\del g / \del x^\lambda})$, and, further,
$p_\mu = J\, p_\mu^{(\alpha )}$.
The equation of motion is thus
\beq
{\del \over \del x^\lambda} ( u^\lambda \, p_\mu ) = 0
\label{48}
\eeq
The energy-momentum tensor corresponding to (\ref{46}) is 
given by
\beq
T_{\mu\nu}= u_{\mu} \, p_{\nu} - \eta_{\mu\nu} (n\, f' - f)
\label{49}
\eeq
where $f' = ({\del f / \del n})$.
The equation of motion (\ref{48}) only coincides with the conservation of $T_{\mu\nu}$ provided $f (n) = n$, so that the
pressure $ P= n \, f' - f = 0$. In this case $p_\mu = n \, u_\mu$.
Thus (\ref{48}) only describes the pressureless flow of a large number of particles.
There is no surprise here, since, beyond using a continuous set to label
the particles, we have not included interparticle interactions which is needed for
nonzero pressure. We can modify $f(n)$ to allow for nonzero pressure, but then
individual particle positions lose meaning as independent dynamical variables.
There are two points to be made here: first the use of individual
positions as dynamical variables to be varied to get equations of motion will not
work, even nonrelativistically. Secondly, having four positions $x^\mu$ was
appropriate for particles because we would eliminate one of them by the
constraint of $H \approx 0$ eventually. We do not have such a constraint
in terms of the fluid variables. Thus we need 3 variables for the fluid velocity whose
variation in the action can lead to the correct equations of motion.
For this, it is useful to recall that, even for the nonrelativistic case, 
an action for fluid dynamics requires the
Clebsch parametrization of the velocities, given by
\beq
v_i = \nabla_i \theta + \alpha \nabla_i \beta
\label{50}
\eeq
The action for nonrelativistic fluids is given by
\beq
S = \int d^4x~ \left[ j^0 \, ( \del_0 \theta + \alpha \, \del_0 \beta )
- {1\over 2} \,j^0\, ( \nabla \theta + \alpha \nabla \beta )^2  - V (j^0 ) \right]
- \int d^4x~ j^0
\label{51}
\eeq
where $j^0 = \rho$ is the density.
Usually the last term can be omitted as it does not contribute to the equations of motion, 
$\int d^3x\, \rho$ being fixed.
Introducing an auxiliary field $j_i$, we can rewrite this as
\beq
S = \int d^4x~ \left[ j^0 \, ( \del_0 \theta + \alpha \, \del_0 \beta ) - j_i 
( \nabla_i \theta + \alpha \nabla_i \beta )
- j^0 + {j_i j_i \over 2 \, j^0} - V (j^0 ) \right]
\label{52}
\eeq
The elimination of $j_i$ evidently leads back to 
(\ref{51}). We notice that this action is the approximation, for $( j^0)^2 \gg j_i j_i$,
of the action
\beqar
S &=& \int d^4x~ \bigl[ j^\mu ( \del_\mu \theta + \alpha \del_\mu \beta ) 
- f (n)\bigr]\nonumber\\
f(n)&=& n +  V(n) , \hskip .3in n^2 = \eta_{\mu\nu} j^\mu \, j^\nu
= (j^0)^2 - j_i j_i
\label{53}
\eeqar
Clearly we can take (\ref{53}) as the relativistic action which reproduces the nonrelativistic one
in the appropriate limit. The point of this argument is that what takes the place of
${\dot x}^\mu$ is the Clebsch parametrization
$\del_\mu \theta + \alpha \del_\mu \beta$.
The distinction between the use of the Clebsch variables and ${\dot x}^\mu$ is not important
if we were just to transform the equations of motion, as Lagrange did; it is relevant only when we
want to construct an action.

With this understanding of the replacement of ${\dot x}_\mu$ by the
Clebsch parametrization, we can now easily generalize the
point-particle action to fluids. Going back to (\ref{6}), we can write the action
\beq
S = \int d^4x~ \left[ 
j^\mu\, (\del_\mu \theta + \alpha \del_\mu \beta ) 
- { i \over 4} j^\mu_{(s)}  \Tr ( \Sigma_3 \,\Lambda^{-1} \del_\mu \Lambda )
- f( n, n_{(s)} )\right]
\label{54}
\eeq
where $\Lambda$ is a function on spacetime, depending on all $x^\mu$ in general.
It is again given explicitly by
$\Lambda = B \, R$, with
\beq
B = {1\over \sqrt{2 n_{(s)} (j_{(s)}^0 + n_{(s)})}} \left[
\begin{matrix} j_{(s)}^0 + n_{(s)} & {\vec\sigma}\cdot{\vec j_{(s)}} \\
{\vec\sigma}\cdot{\vec j_{(s)}}  & j_{(s)}^0 + n_{(s)} \\
\end{matrix}
\right]
\label{55}
\eeq
There are two currents in (\ref{54}), $j^\mu$ for mass and
$j^\mu_{(s)}$ for spin. As before, $n^2 = \eta_{\mu\nu} j^\mu j^\nu$,
$n_{(s)}^2 = \eta_{\mu\nu} j_{(s)}^\mu j_{(s)}^\nu$.
For point particles, these two currents were proportional to each other,
rather the corresponding momenta were the same.
However, for fluids, we can consider independent transport of
mass and spin, so the general situation is
to have separate currents. 

So far we have considered only the mass and spin of the fluid. The inclusion of additional quantum numbers is straightforward. We consider the full symmetry group,
which is of the form of the Poincar\'e group times the internal symmetry group $G$.
The latter could be the full symmetry group of the standard model, for example, including the gauged subgroup $U(1)_Y \times SU(2)_L \times SU(3)_c$, as well as the relevant chiral symmetries, the difference between the baryon and lepton numbers ($B-L$), etc.
The generalization of the action (\ref{54}) is then
\beqar
S &=& \int d^4x~ \Bigl[ 
j^\mu\, (\del_\mu \theta + \alpha \del_\mu \beta ) 
- { i \over 2} \, j^\mu_{(s)}  \Tr ( \Sigma_3 \,\Lambda^{-1} \del_\mu \Lambda )
+ i \sum_a\, j^\mu_{(a)} \Tr (h_a \, g^{-1} D_\mu \, g) \nonumber\\
&&\hskip .7in
- f( \{ n \} ) )\Bigr] ~+~ S(A)
\label{56}
\eeqar
where we have, in principle, separate currents $j^\mu, \, j^\mu_{(s)}, \, j^\mu_{(a)}$
for all the diagonal generators
of the symmetry group; i.e., as many currents as the rank of the group.
We use $\theta$, $\alpha$, $\beta$ to describe mass transport, $\Lambda$ for spin transport, 
and $g \in G$ for the transport of internal symmetries corresponding to a group $G$.
(We use a current of mass dimension $3$, so that $j^\mu$ is like
the mass current divided by a mass parameter.)
The covariant derivative $D_\mu$ indicates gauging with respect to the
gauge fields of the standard model. $S(A)$ indicates the part of the action for the
gauge fields, which is the sum of Yang-Mills-type terms.
The function $f( \{ n \} )$ now depends on all the invariants
of the form $n = \sqrt{j^\mu \, j^\nu \, \eta_{\mu\nu}}$ made from the currents, including, in principle, terms of the form $\sqrt{j^\mu \, j_{(s)}^\nu \, \eta_{\mu\nu}}$, $\sqrt{j^\mu \, j_{(a)}^\nu \, \eta_{\mu\nu}}$, etc.

The action (\ref{56}) is a general action for relativistic (Abelian/nonabelian) 
magnetohydrodynamics.
The distinction between different types of fluids with the same symmetry
is in the choice of the function $f$ (which determines the various partial
pressures and hence the equation of state) and ``constitutive relations"
among the currents. For example, if we have only one species of particles,
each carrying an electric charge $e$, then for the corresponding mass current
and electric current we expect the relation
$j_{(e)}^\mu  = e\, j^\mu$.

\section{Magnetohydrodynamics with spin}

Most of the terms in the action (\ref{56}) were already given many years ago
in \cite{bistro}, see also \cite{review}.
We have also considered a similar action with Wess-Zumino terms added
\cite{{NRR},{CNT}}
to account for anomalies and have shown that the chiral magnetic effect and the chiral
vorticity effect are incorporated.
The main new point here is the clarification of how the terms associated with
the Poincar\'e symmetry enter. To illustrate how such terms can 
affect the physics of the fluid, we will consider in some detail a special case of 
(\ref{56}) where we take the flow velocity for
 the mass, spin and electric charge to be the same, i.e.,
 $j^\mu_{(s)} = s j^\mu$, $j_{(e)}^\mu  = e\, j^\mu$.
 If we have a single species of particles with identical charge we should expect the same velocity
 for mass and charge. Even so, the spin flow can have a different velocity as spin singlet
 combinations can form; their transport would affect mass/charge flow but not the spin current.
 However, for a dilute system where such combinations are unlikely on the scale of coarse-graining, having the same flow velocity for spin as well is not unreasonable.
 This is essentially this special case; we 
will analyze this in some more detail as it is closely related to
 the single-particle motion discussed in earlier sections.
 Thus we consider the action
\beq
S =  S(A) + \int d^4x~
 \Bigl[j^\mu\, (\del_\mu \theta + \alpha \del_\mu \beta + e A_\mu ) 
- i \,{ s \over 2} \, j^\mu\, \Tr ( \Sigma_3 \,\Lambda^{-1} \del_\mu \Lambda )
- f( n, \sigma ) \Bigr]
\label{57}
\eeq
Anticipating that we will need magnetic moment couplings, we take
$f$ to be a function of
$n = \sqrt{j^\mu j_\mu }$ and $\sigma = S^{\mu\nu} F_{\mu\nu}$.
For obtaining and simplifying the equations of motion, it is convenient to use
a slightly different action
 \beqar
S &=& \int d^4x~
\Bigl[j^\mu\,(\del_\mu \theta + \alpha \del_\mu \beta + e A_\mu ) 
- { i s \over 2}  j^\mu\, \Tr ( \Sigma_3 \,\Lambda^{-1} (\del_\mu + i\, \gamma_{\mu\nu}
\xi^\nu ) \Lambda )\Bigr]\nonumber\\
&&\hskip .3in
-\int d^4x~ f( n, \sigma ) ~+~ S(A)
\label{58}
\eeqar
In (\ref{58}), we treat $\Lambda$ as an arbitrary dynamical variable. 
The requirement
that $\Lambda$ can be written as $B\,R$ with the velocity in $B$
being identical to the mass/charge transport velocity $u^\mu={{j^\mu} \over {\sqrt{j^2}}}$ is enforced by
the constraint 
\beq
S^{\mu \nu} \, u_\nu = 0
\label{59}
\eeq
This is
obtained as the equation of motion of the Lagrange multiplier field $\xi^\nu$.
First we show that this constraint can indeed give the identity of the flow velocities.
Writing out (\ref{59}),
we find
\beq
\Tr [ \Lambda^{-1} \, \gamma\cdot u \Lambda , \Sigma_3 ] \, \Lambda^{-1} \gamma_\nu
\Lambda  = \Lambda_\nu ^{~\alpha} ~
\Tr [ \Lambda^{-1} \, \gamma\cdot u \Lambda , \Sigma_3 ] \, \gamma_\alpha
= 0
\label{60}
\eeq
Since $\Lambda_\nu ^{~\alpha}$ is invertible, we need 
$ [ \Lambda^{-1} \, \gamma\cdot u \Lambda , \Sigma_3 ]$ to have zero trace with $\gamma_\alpha$, for any $\alpha$. Further, since $ [ \Lambda^{-1} \, \gamma\cdot u \Lambda , \Sigma_3 ]$ is a linear combination of single powers of $\gamma_\mu$, we get
$ [ \Lambda^{-1} \, \gamma\cdot u \Lambda , \Sigma_3 ] = 0$ or
\beq
\Lambda^{-1} \, \gamma\cdot u\,\, \Lambda  = a \, \gamma_0 + b \, \gamma_3
\label{61}
\eeq
Since $u^\mu u_\mu =1$, $a = \cosh \omega$, $b = \sinh \omega$, so that we may write
$ \Lambda^{-1} \, \gamma\cdot u \Lambda = \Lambda_0^{-1} \gamma_0\, \Lambda_0$,
$\Lambda_0 = \exp ( -i \,\omega\, \gamma_{03})$.
The solution for $\Lambda$ is thus
$\Lambda = B\, R\, \Lambda_0$, with $B$ as given in 
(\ref{55}). $\Lambda_0$ drops out of the action
since it commutes with $\Sigma_3$ and $\Tr  (\Sigma_3 \, \gamma_{03}) = 0$.
We may thus drop it from further consideration.
Thus the constraint (\ref{59}) does enforce the equality of the flow velocities.

The remaining variational equations are
\beq
\del_\mu \, j^\mu = 0, \hskip .2in
j^\mu \, \del_\mu \alpha = 0, \hskip .2in
j^\mu \, \del_\mu \beta = 0
\label{62}
\eeq
\beqar
V_\mu - K_\mu + S_{\mu\nu} \xi^\nu &=& {\del f \over \del n} u_\mu
\label{63}\\
V_\mu &=& \del_\mu \theta + \alpha\, \del_\mu \beta + e A_\mu \nonumber\\
K_\mu &=&  { i s \over 2}\,\Tr ( \Sigma_3 \,\Lambda^{-1} \del_\mu  \Lambda )
\nonumber
\eeqar
\beqar
\D S_{\mu\nu} - {2\over n} \,{\del f \over \del \sigma}\, 
(S_{\mu\lambda} F^\lambda_ {~\, \nu} - S_{\nu \lambda} F^{\lambda}_{~\, \mu})
+ ( u_\mu S_{\nu \lambda} - u_\nu S_{\mu \lambda} ) \xi^\lambda 
&=&0
\label{64}
\eeqar
where $\D = u^\mu \del_\mu$.
The first set (\ref{62}) arises from varying the action with respect to
$\theta$, $\alpha$ and $\beta$. The second one, (\ref{63}),
corresponds to the variation with respect to $j^\mu$ and the last one is due to
the variation of $\Lambda$ as in $\delta \Lambda = -i  \omega^{\mu\nu} \gamma_{\mu\nu} \, \Lambda$.

The simplification of these equations will proceed as in the spinless case.
It is convenient in what follows to denote $\gamma_{\mu\nu}$ by $t_A$, taking
$A,\, B, \, C$ as composite indices; thus $t_{12} = \half \Sigma_3$.
The normalization is $\Tr (t_A t_B) = \delta_{AB}$.
We can then write
\beq
\Lambda^{-1} \, d \Lambda = -i t^A \, {\cal E}_{A},
\hskip .3in
\Lambda^{-1} \, \del_\mu \Lambda = -i t^A \, {\cal E}_{A, M}\, \del_\mu \vf^M
\label{65}
\eeq
where we denote the parameters of $\Lambda$ generically by $\vf^M$.
This relation gives
\beq
K_\mu = s \, {\cal E}_{12, M} \,\del_\mu \vf^M
\label{66}
\eeq
Further, from the definition of $S^A = {s \over 2} \Tr ( \Sigma_3 \, \Lambda^{-1} t^A \Lambda )$,
we find
\beqar
f_{ABC} S^A \, dS^B\, \wedge\, dS^C &=& s^3 f_{12,MP} f^{12,MN} f^{12,PQ}
\, {\cal E}_N \,\wedge\,  {\cal E}_Q\nonumber\\
&=& s^3 f^{12, NQ} \ {\cal E}_N \,\wedge\,  {\cal E}_Q
\label{67}
\eeqar
where we used the relation
$f_{12,MP} f^{12,MN} f^{12,PQ} = f^{12, NQ}$ which may be shown directly from the definitions.
Using this equation and taking the curl of $K_\mu$, we find
\beqar
\del_\mu K_\nu - \del_\nu K_\mu &=& - {i s \over 2} \Tr \Sigma_3 [ \Lambda^{-1}\del_\mu \Lambda ,
\Lambda^{-1} \del_\nu \Lambda ] \nonumber\\
&=& - {1 \over s^2} f_{ABC} S^A \,\del_\mu S^B \, \del_\nu S^C\nonumber\\
&=& { 4 \over s^2} \, S_{\alpha\beta} \,
\del_\mu S^{\lambda \alpha} \, \del_\nu S_\lambda^{~\,\beta}
\label{68}
\eeqar

Taking the curl of (\ref{63}) and contracting with $u^\mu$ and using (\ref{62}) and
(\ref{68}), we get
\beq
\D (f' u_\nu) - \del_\nu f' = e \,u^\lambda \, F_{\lambda \nu}  - { 4 \over s^2} \,  
S_{\alpha\beta} \,
\D S^{\lambda \alpha} \, \del_\nu S_\lambda^{~\,\beta}
+ \D ( S_{\nu \lambda}\xi^\lambda ) -
u^\mu \del_\nu (S_{\mu \lambda} \xi^\lambda )
\label{69}
\eeq
where $f' = ({\del f / \del n})$.
The equation  for the spin density, namely (\ref{64}), remains as it is for now. The key issue however
is that the constraint (\ref{59}) must be preserved by the evolution equations.
The requirement is that
\beq
\D ( S_{\mu \nu} u^\nu ) = 0 
\label{70}
\eeq
on the constrained subspace with $S_{\mu \nu} u^\nu =0 $.
Using (\ref{64}) and (\ref{69}), this can be written as
\beq
S_{\mu\nu} \Bigl[  u_\lambda F^{\lambda \nu} X(n,\sigma)+ \D (S^{\nu\lambda} \xi_\lambda ) - u^\rho \del^\nu ( S_{\rho \lambda} \xi^\lambda )
- {4 \over s^2} \, S_{\alpha\beta} \, (\D S^{\lambda \alpha}) \, \del^\nu S_\lambda^{~\beta}
+ \del^\nu f' + f' \xi^\nu ~\Bigr]= 0
\label{71}
\eeq
where
\beq
X(n, \sigma) = e- {2\over n} {\del f \over \del n} {\del f \over \del \sigma} 
\label{71a}
\eeq
The vector inside the bracket in (\ref{71}) is orthogonal to $S_{\mu\nu}$, so it can be written as a linear combination of $u^{\nu}$ and $W^{\nu}$, where $W^{\nu}$ is the normalized Pauli-Lubanski vector defined as
\beq
W^{\mu} = -{1 \over {2s}} \epsilon^{\mu\nu\alpha\beta} u_{\nu} S_{\alpha\beta} ~~~,~~~~~~~W\cdot W = -1 
\label{71b}
\eeq
The following relations, which are easily verified, are useful for further simplification,
\beqar
S_{\mu\nu} & =& s \epsilon _{\mu \nu \rho \sigma} W^{\rho}u^{\sigma} ~~~~~\Longrightarrow S_{\mu\nu} u^{\nu} = S_{\mu\nu} W^{\nu}= W \cdot u =0 
\label{71c}\\
S_{\mu\nu}S^{\nu\rho} &=& -s^2  ( \delta^{\rho}_{\mu} + W_{\mu}W^{\rho} - u_{\mu}u^{\rho})
\label{71d}
\eeqar
By virtue of these relations,
a vector $\omega^{\nu}$, satisfying $S_{\mu\nu}\omega^{\nu} =0 $ 
is of the form
 $\omega^{\mu} = u^{\mu} (u \cdot \omega) - W^{\mu} (W \cdot \omega) $. Applying this to (\ref{71}) we get
\beqar
&&X(n, \sigma) [u_\lambda F^{\lambda \nu} + W^{\nu} W_{\rho} u_{\lambda} F^{\lambda\rho}] + \D (S_{\nu\lambda} \xi^\lambda ) - u_\rho \del^\nu ( S^{\rho \lambda} \xi_\lambda )\nonumber \\
&&+ ~W^{\nu} [ W_{\rho} \D(S^{\rho\lambda} \xi_\lambda )-u_{\rho} W_{\sigma} \del^{\sigma} (S^{\nu\lambda} \xi_\lambda ) ] - {4 \over s^2} \, S_{\alpha\beta} \, (\D S^{\lambda \alpha})\, (\del^\nu S_\lambda^{~\beta} + W^{\nu} W_{\sigma} \del^{\sigma} S_{\lambda}^{\beta} ) \nonumber \\
&&+ ~\del^\nu f' + W^{\nu}W_{\rho} \del^{\rho} f'- u^\nu \D f' + f' \xi^\nu  = 0
\label{73}
\eeqar
Equation (\ref{73}) is to be understood as the equation determining $\xi_{\nu}$. Further,  from the way $\xi_{\nu}$ it enters the action (\ref{58}),
it is clear that, without loss of generality,
 we can take $\xi^\nu$ to be orthogonal to $u_{\nu}$ and $W_{\nu}$ as well. We can solve (\ref{73}) as a series expansion in gradient terms and powers of the external field $F$, namely
 \beq
 \xi^\nu = \xi^{\nu}_{ (1)} ~+~  \xi^{\nu}_ {(2)} + \cdots
 \label{73a}
 \eeq
 where $\xi^{\nu}_{ (1)}$ contains terms linear in gradients or in $F$; $\xi^{\nu}_{ (2)}$ contains terms quadratic in gradients or quadratic in $F$ or linear in gradients and $F$ and so on. In this
 way,  we find
\beqar
 \xi^{\nu}_{(1)} &=& {1\over f'} \left[ u^\nu \, \D f' - \del^\nu f' -W^{\nu}W_{\rho} \del^{\rho} f'-X(n, \sigma) [u_\lambda F^{\lambda \nu} + W^{\nu} W_{\rho} u_{\lambda} F^{\lambda\rho}] \right]\label{73b} \\
  \xi^{\nu}_{(2)} &=& -{1 \over f'} \Bigl[ \D (S^{\nu\lambda} \xi_{\lambda(1)} ) - u_\rho \del^\nu ( S^{\rho \lambda} \xi_{\lambda (1)})+ ~W^{\nu} [ W_{\rho} \D(S^{\rho\lambda} \xi_{\lambda (1)} )-u_{\rho} W_{\sigma} \del^{\sigma} (S^{\rho\lambda} \xi_{\lambda (1)} ) ] \nonumber \\
  &&- {4 \over s^2} \, S_{\alpha\beta} \, (\D S^{\lambda \alpha}_{(1)})\, (\del^\nu S_\lambda^{~\beta} + W^{\nu} W_{\sigma} \del^{\sigma} S_{\lambda}^{\beta} ) \Bigr]
  \label{73c}
  \eeqar
We can now substitute these expressions into (\ref{64}) and (\ref{69}) to obtain the equations
of motion for the spin density $S_{\mu\nu}$  and the fluid velocity $u_\nu$.
The expressions are quite involved, so, at
this point we will simplify the equations by imposing the condition
  \beq
  X(n,\sigma) =0~~~~\Longrightarrow ~~~e= {2\over n} {\del f \over \del n} {\del f \over \del \sigma} 
\label{72}
\eeq
This determines the dependence of $f$ on $\sigma$ in terms of its dependence
on $n$. As discussed earlier, in the absence of the external field $F$, we take $f(n) = n +V(n)$. ($V(n)=0$ corresponds to the pressureless case.) One can find a power series solution for $f(n, \sigma)$, satisfying (\ref{72}), of the form
\beq
f(n, \sigma) = n +V(n) + {{en \sigma} \over {2[1+ V'(n)]}} + {\cal{O}} (\sigma^2)
\label{72b}
\eeq
This can be thought of as the analog of the requirement of $g = 2$ in the single particle case. 
In the nonrelativistic limit, $V'(n) \ll 1$ and we see from the
term linear in $\sigma$ that the magnetic moment is proportional to the charge density $en$ and corresponds to $g=2$.
The condition (\ref{72}) characterizes a special kind of fluid,
perhaps the most relevant case, since $g-2$ is usually very small.
Nevertheless, we emphasize that this is a specialization; one can always use the
general solution (\ref{73b}), (\ref{73c}) for more general types of fluids.

Using (\ref{73b}-\ref{73c}) in (\ref{64}) and (\ref{69}) and keeping only terms linear in $F$ and gradients we find the following expressions for the equations of motion
\beqar
\D( f' \, u_\nu ) - \del_\nu f' &=&
e \Bigl[ u^\lambda \, F_{\lambda \nu} - 
{4\over {s^2 f'}}\, \del_\nu S^{\lambda \beta} ( S F S - F S S )_{\lambda \beta} - {1 \over {f'^2}} (S_{\nu\alpha} F^\alpha_\lambda- S_{\lambda\alpha} F^\alpha_\nu) \del^{\lambda} f' \Bigr] \label{76} \nonumber\\
&&\hskip .2in+ \cdots
\\
\D S_{\mu\nu} &=& {1\over f' } \left[ S_{\mu}^{~\lambda} (e F_{\lambda \nu} + G_{\lambda \nu} )
- S_{\nu}^{~\lambda} (e F_{\lambda \mu} + G_{\lambda \mu} )\right] \nonumber \\
&&\hskip .2in - {{4e} \over {s^2 f^{' 2}}} ( u_\mu S_{\nu}^{~\lambda} - u_\nu S_{\mu}^{~\lambda} )
\del_\lambda S^{\rho \beta} ( S F S - F S S)_{\rho \beta} \nonumber \\
&&\hskip .2in + {e \over {f'^3}} \Bigl[ s^2 \left[(u_\mu F_\nu^\lambda -u_\nu F_\mu^\lambda) + (u_\mu W_\nu -u_\mu W_\nu)W^{\rho}F_\rho^\lambda \right] \nonumber \\
&&\hskip .2in + u_\mu (SFS)_{\nu\lambda} -u_\nu (SFS)_{\mu\lambda} \Bigr] \del_{\lambda} f' 
+ \cdots
\label{77}\\
G_{\lambda \nu} &=& u_\lambda \, \del_\nu f' - u_\nu \, \del_\lambda f'
\nonumber\\
(S F S - F S S )_{\lambda \beta} 
&=& S_{\lambda}^{~\rho}\, F_{\rho \tau}\, S^{\tau}_{~\beta} -
F_{\lambda}^{~\rho} \, S_{\rho \tau} \, S^{\tau}_{~\beta} 
\label{78}
\eeqar
We have ignored the gradients of the external field, so that this is valid for
almost uniform fields, or as the first set of terms in an expansion in terms
of gradients of the fields.

The first equation in this set, (\ref{76}), is the analog of the Euler equation with the
Lorentz force on the right hand side, as expected for magnetohydrodynamics.
(The term magnetohydrodynamics is often used for the more restricted case
where the electric field ${\vec E}$ is related to the magnetic field
${\vec B}$ via ${\vec E}  + {\vec v} \times  {\vec B} = 0$, which plays the role of Ohm's law
for a plasma.
We are using the term in a more general sense. The specialization
to ${\vec B}$ via ${\vec E}  = {\vec v} \times  {\vec B}$ can be easily made at any stage.)
The appearance of a term involving the gradient of the spin density
on the right hand side of this equation is not surprising since a 
term like $S^{\mu\nu} F_{\mu\nu}$ in the Hamiltonian would be like a contribution to
the potential energy and we should expect its gradient in the equation of motion.

The second equation (\ref{77}) describes the flow (or precession)
of the spin density.
What is novel and interesting is that this equation
 shows a precession term
$S_\mu^{~\lambda} G_{\lambda \nu} - S_\nu^{~\lambda} G_{\lambda \mu} $ in addition to the
usual precession effect due to  $e (S_\mu^{~\lambda} F_{\lambda \nu} - S_\nu^{~\lambda} F_{\lambda \mu} )$, even in the absence of gradients for $S^{\alpha\beta}$.
Since $G_{\lambda \nu}$ involves gradients of the pressure and energy density, 
we see that nonuniform pressure and energy density
in magnetohydrodynamics can generate precessional motion for spin density.
Notice that we may rewrite the Euler equation (\ref{76}) also as
\beq
f' \, \D u_\nu = 
 \left[ u^\lambda \, (e \,F_{\lambda \nu}  + G_{\lambda \nu} ) - 
{4\over{s^2 f'}}\, \del_\nu S^{\lambda \beta} ( S F S - F S S )_{\lambda \beta}
\right] ~+~ \cdots\label{79}
\eeq
Thus $G_{\lambda \nu}$ plays a role similar to that of $F_{\lambda \nu}$ in this equation, so its presence in
the spin precession equation (\ref{77}) is not entirely surprising.

\section{Discussion}

The group-theoretic formulation of fluid dynamics has been used to describe fluids with nonabelian charges and to include anomaly effects. We explore this formulation further in this paper.
The focus here is to clarify the role of Poincar\'e group, rather than internal symmetries.
For this, we started by considering relativistic charged particle dynamics in some detail.
The minimal symplectic form, as given by the Poincar\'e group, along with the Lorentz force
is shown to imply a gyromagnetic ratio of $2$. (We give a clearer definition of what is meant by minimal in the text.)
A similar result was found some time ago
in $2+1$ dimensions. In that case, it is known that variants of the symplectic form
can accommodate anomalous magnetic moment. We show that a similar result holds in
$3+1$ dimensions as well.
We analyze the canonical structure and also show how the one-particle wave equation with the correct magnetic moment and spin-orbit interactions can be obtained upon quantization.

The extension to fluids is then considered and
the general group-theoretic framework is clarified further.
The main result may be summarized as equation
(\ref{56}) which gives the general form of the action for 
a fluid with Poincar\'e symmetry and an internal symmetry corresponding to group $G$.
This action with the addition of a suitable Wess-Zumino term for anomalies
should describe general fluid dynamics with anomalous symmetries as well.
The derivatives can also be made Levi-Civita covariant
to accommodate gravitational effects.
Variants of (\ref{56}) have been used previously, with and without anomalies, to describe
a number of phenomena \cite{{bistro},{NRR}, {CNT}, {capasso},{dai-nair}}.

We also considered another special case, namely,
 the extension of standard magnetohydrodynamics
(Maxwell field coupled to charged fluids) to include spin effects. The nature of this theory
 is dictated purely on symmetry grounds by the Poincar\'e or Lorentz group.
 The equation for the fluid shows new spin precession effects due to
 the gradients of pressure and energy density. There are also
 corrections to the Euler equation depending on the gradients of the spin density
 in the presence of electric and magnetic fields.

\bigskip

This research was supported in part by National Science Foundation grants PHY-1068172  and PHY-1213380 
and by PSC-CUNY awards.

\end{document}